\documentclass[prb,twocolumn,aps,superscriptaddress,showpacs,floatfix]{revtex4}
\usepackage{amsmath}
\usepackage{graphicx}
\usepackage{color}
\usepackage{tikz}
\usepackage{subfigure}
\begin{document}
\title{
Generation of localized magnetic moments in the charge-density-wave state
}

\author{R.S. Akzyanov}
\affiliation{Moscow Institute for Physics and Technology (State
University), 141700 Moscow Region, Russia}
\affiliation{Institute for Theoretical and Applied Electrodynamics, Russian
Academy of Sciences, 125412 Moscow, Russia}
\affiliation{All-Russia Research Institute of Automatics, Moscow, 127055 Russia }
\author{A.V. Rozhkov}
\affiliation{Moscow Institute for Physics and Technology (State
University), 141700 Moscow Region, Russia}
\affiliation{Institute for Theoretical and Applied Electrodynamics, Russian
Academy of Sciences, 125412 Moscow, Russia}

\begin{abstract}
We propose a mechanism explaining the generation of localized magnetic
moments in charge-density-wave compounds. Our model Hamiltonian describes
an Anderson impurity placed in a host material exhibiting the
charge-density wave. There is a region of the model's parameter space,
where even weak Coulomb repulsion on the impurity site is able to localize
the magnetic moment on the impurity. The phase diagram of a single impurity
at
$T=0$
is mapped. To establish the connection with experiment thermodynamic
properties of a random impurity ensemble is studied. Magnetic
susceptibility of the ensemble diverges at low temperature; heat capacity
as a function of the magnetic field demonstrates pronounced low field peak.
Both features are consistent with experiments on orthorhombic
TaS$_3$
and blue bronze.
\end{abstract}


\maketitle

\section{Introduction}

Study of coexistence and competition between different types of order is a
recurrent theme of the modern condensed matter research, both theoretical
and experimental. In this paper we discuss a particular example of such
coexistence. There is significant experimental evidence that several
charge-density wave (CDW) materials
(orthorhombic TaS$_3$,
Ref.~\onlinecite{biljakovic_otas3_epl,biljakovic_otas3_others};
blue bronze
Ref.~\onlinecite{biljakovic_bbrz};
non-magnetic tritellurides YTe$_3$ and LaTe$_3$,
Ref.~\onlinecite{nancy_ru_thesis})
show unusual sensitivity to the magnetic field at low temperatures.
For example, magnetic susceptibility of o-TaS$_3$ is
temperature-independent in a broad range of temperatures,
both above and below its CDW transition temperature
$T_{\rm CDW}$=218\,K
(see Fig.1 of
Ref.~\onlinecite{biljakovic_otas3_epl}).
However, below
$\sim$60\,K
the susceptibility begins to grow quickly as the temperature decreases.
Experimental investigation of low-temperature magnetic and thermodynamic
properties
concluded~\cite{biljakovic_otas3_epl}
that such a behavior is consistent with the assumption that a disordered
ensemble of localized magnetic moments undergoes a transition into a glass
state.

This magnetic glass is quite unexpected (non-magnetic
glass~\cite{bragg_glass}
induced by the CDW pinning is, of course, possible, but it has little
relevance to the issue under consideration). Theoretically, the CDW
magnetic properties are believed to be trivial: due to the gap in the
electron spectrum, at temperatures significantly below
$T_{\rm CDW}$
electronic contribution to the susceptibility vanishes. How localized
magnetic moments can emerge under such circumstances?

In this paper we propose a mechanism, which may explain the origin of these
magnetic moments. We will study an impurity inside a CDW compound. Assuming
weak electron repulsion at the impurity site, we find a parameter regime
where exactly one electron resides on that site. Spins of such electrons
are responsible for low-temperature magnetism of the CDW material.

Our mechanism is quite generic, and does not impose special restrictions on
the dimensionality of the Hamiltonian, band structure, and other model
details. However, the low-temperature magnetism of an CDW compound is by no
means a universal feature: if the system parameters are outside of the
required range, the magnetic response is completely trivial. Of course, a
particular material may enter this regime either by luck, or by intelligent
design of a material scientist.

Our paper is organized as follows. In
Sec.~\ref{sect::model}
we introduce our model Hamiltonian. Generation of the magnetic moment on a
single impurity is discussed in
Sec.~\ref{sect::generation}.
Thermodynamic properties of an ensemble of these impurities is investigated
in
Sec.~\ref{sect::ensamble}.
The discussion are presented in
Sec.~\ref{sect::discussion}.
We conclude in
Sec.~\ref{sect::conclusions}.

\section{Model}
%%%%%%%%%%%%%%%%%%%%%%%%%%%%%%%%%%%%%%%%%%%%%%%%%%
\label{sect::model}
%%%%%%%%%%%%%%%%%%%%%%%%%%%%%%%%%%%%%%%%%%%%%%%%%%

We study a single Anderson impurity, which is located inside the host
material with CDW ground state. The model's Hamiltonian $H$ is equal to
\begin{eqnarray}
%%%%%%%%%%%%%%%%%%%%%%%%%%%%%%%%%%%%%%%%%%%%%%%%%%
\label{H}
%%%%%%%%%%%%%%%%%%%%%%%%%%%%%%%%%%%%%%%%%%%%%%%%%%
H=H_{\rm F} + H_{\rm imp} + H_{\rm hop},
\end{eqnarray}
where Fr\"ohlich Hamiltonian
$H_{\rm F}$,
Anderson Hamiltonian
$H_{\rm imp}$,
and hybridization Hamiltonian
$H_{\rm hop}$
are defined as follows:
\begin{eqnarray}
H_{\rm F}=\sum \limits_{\mathbf{k}\sigma} \left(\epsilon_{\mathbf{k}}
a_{1,\mathbf{k}\sigma}^{\dagger} a_{1,\mathbf{k}\sigma} -
\epsilon_{\mathbf{k}} a_{2,\mathbf{k}\sigma}^{\dagger}
a_{2,\mathbf{k}\sigma}\right) +
\nonumber
\\
%%%%%%%%%%%%%%%%%%%%%%%%%%%%%%%%%%%%%%%%%%%%%%%%%%
\label{H_F}
%%%%%%%%%%%%%%%%%%%%%%%%%%%%%%%%%%%%%%%%%%%%%%%%%%
\sum \limits_{\mathbf{k}\sigma}
\Delta\left(
	e^{i\phi}
	a_{1,\mathbf{k}\sigma}^{\dagger}
	a_{2,\mathbf{k}\sigma}
	+
	\text{H.c.}
	\right),
\\
H_{\rm imp} =\sum \limits_{\sigma} (-\epsilon_{0})
d_{\sigma}^{\dagger}d_{\sigma} +
Ud^{\dagger}_{\uparrow}d_{\uparrow}d^{\dagger}_{\downarrow}d_{\downarrow} ,
\\
%%%%%%%%%%%%%%%%%%%%%%%%%%%%%%%%%%%%%%%%%%%%%%%%%%
\label{H_hop}
%%%%%%%%%%%%%%%%%%%%%%%%%%%%%%%%%%%%%%%%%%%%%%%%%%
H_{\rm hop}
=
\sum \limits_{\mathbf{k}\sigma}
t(a_{1,\mathbf{k}\sigma}^{\dagger}+a_{2,\mathbf{k}\sigma}^{\dagger})
d_{\sigma}+\text{H.c.}
\end{eqnarray}
This model describes two bands of electrons with perfect nesting. The
dispersion of the first band is
$\epsilon_{\mathbf{k}}$,
the dispersion in the second band is
$-\epsilon_{\mathbf{k}}$.
Band electron creation operators are
$a^\dag_{\alpha,\mathbf{k}\sigma}$,
where
$\alpha=1,2$,
is the band index, $\mathbf{k}$ and $\sigma$ are momentum and spin of the
electron.

The CDW phase is characterized by a finite value of the order parameter
$\Delta$,
which we assume to be real and positive. The quantity $\phi$ in
Eq.~(\ref{H_F})
equals to CDW phase
\begin{eqnarray}
%%%%%%%%%%%%%%%%%%%%%%%%%%%%%%%%%%%%%%%%%%%%%%%%%%
\label{phi_vs_Rimp}
%%%%%%%%%%%%%%%%%%%%%%%%%%%%%%%%%%%%%%%%%%%%%%%%%%
\phi= \mathbf{K}_{\rm n} \cdot \mathbf{R}_{\rm imp}
\end{eqnarray}
at the location of the impurity
${\bf R}_{\rm imp}$. (In this equation
${\bf K}_{\rm n}$
is the nesting vector of the Fermi surface.)

Operator
$d^\dag_\sigma$
creates an electron at the impurity site. Single-electron energy at the
impurity is equal to
$-\epsilon_0$.
The hybridization amplitude between the impurity state and the bands is
$t$. The electron-electron interaction at the impurity site is $U>0$. In
this paper we assume that $U$ is small:
\begin{eqnarray}
%%%%%%%%%%%%%%%%%%%%%%%%%%%%%%%%%%%%%%%%%%%%%%%%%%
\label{small_U}
%%%%%%%%%%%%%%%%%%%%%%%%%%%%%%%%%%%%%%%%%%%%%%%%%%
U \ll \Delta.
\end{eqnarray}
This is a necessary condition for the use of the perturbation theory in
powers of the interaction.

\section{Single-impurity properties}
%%%%%%%%%%%%%%%%%%%%%%%%%%%%%%%%%%%%%%%%%%%%%%%%%%
\label{sect::generation}
%%%%%%%%%%%%%%%%%%%%%%%%%%%%%%%%%%%%%%%%%%%%%%%%%%

%\subsection{The impurity inside metallic host ($T > T_{\rm CDW}$)}

First, let us briefly discuss the properties of our model in the ``high
temperature" regime,
$T > T_{\rm CDW}$,
$\Delta (T) = 0$.
When $U$ is small, and the temperature is high, the Kondo correlations at
the impurity site are negligible: combining
inequality~(\ref{small_U})
and inequality
$T_{\rm K} \ll U$,
we derive
\begin{eqnarray}
T_{\rm K} \ll U \ll \Delta(0) \sim T_{\rm CDW} < T
\end{eqnarray}
for this regime. Consequently, the interaction at the impurity site can be
treated perturbatively. Thus, above the transition into the ordered state
our model describes electrons experiencing potential scattering off the
impurity. When we generalize our model to include many impurities randomly
placed in the sample, we should recover the usual phenomenology of a metal
with disorder.

%\subsection{Generation of magnetic moment}
The low temperature behavior of the model is much less trivial, as we will
see below. Experimentally, the magnetic susceptibility starts to diverge
when the temperature is significantly smaller than
$T_\text{CDW}$.
Thus, we will study the regime
\begin{eqnarray}
%%%%%%%%%%%%%%%%%%%%%%%%%%%%%%%%%%%%%%%%%%%%%%%%%%
\label{low_T}
%%%%%%%%%%%%%%%%%%%%%%%%%%%%%%%%%%%%%%%%%%%%%%%%%%
T \ll T_\text{CDW}.
\end{eqnarray}
In this limit the CDW order parameter is independent of temperature, and it
is permissible to use its zero-temperature value for calculations.

For weak interaction,
Eq.~(\ref{small_U}),
perturbation theory is allowed. To apply the perturbation theory we need to
find the eigenstates and eigenenergies of the unperturbed Hamiltonian. The
unperturbed Hamiltonian is given by
Eq.~(\ref{H})
in which we put
$U=0$.
%\begin{eqnarray}\label{H_0}
%\hat{H_0}=\sum \limits_{k} (\epsilon_{k} a_{1,k}^{\dagger} a_{1,k} -
%\epsilon_{k} a_{2,k}^{\dagger} a_{2,k} \nonumber
% \\
% + |\Delta|e^{i\phi}a_{1,k}^{\dagger} a_{2,k}+|\Delta|e^{-i\phi}a_{2,k}^{\dagger} a_{1,k}-\epsilon_{0} d_{}^{\dagger}d_{}
% \nonumber
% \\
% +t(a_{1,k}^{\dagger}+a_{2,k}^{\dagger})d_{}+td_{}^{\dagger}(a_{1,k}+a_{2,k}))
%\end{eqnarray}
It can be diagonalized straightforwardly. It is convenient to introduce a set
of creation operators corresponding to eigenstates with energy $E$
\begin{eqnarray}
A_E
=
\delta_E d^{\dagger}
+
\sum \limits_{\mathbf{k}}
	\beta_{E\mathbf{k}}
a_{1,\mathbf{k}}^{\dagger}+\gamma_{E\mathbf{k}} a_{2,\mathbf{k}}^{\dagger},
\end{eqnarray}
where
$A_E$
satisfies the equation:
\begin{eqnarray}
%%%%%%%%%%%%%%%%%%%%%%%%%%%%%%%%%%%%%%%%%%%%%%%%%%
\label{A_eq}
%%%%%%%%%%%%%%%%%%%%%%%%%%%%%%%%%%%%%%%%%%%%%%%%%%
[H,A_E]=EA_E.
\end{eqnarray}
Here
$\delta_{E}$,
$\beta_{E\mathbf{k}}$,
and
$\gamma_{E\mathbf{k}}$
are $c$-number coefficients. We do not show the spin index explicitly since
the
$U=0$
Hamiltonian can be split into two identical Hamiltonians for two spin
projections.

\subsection{Subgap bound state}

Equation~(\ref{A_eq})
is equivalent to a system of linear equations on coefficients
$\delta_E$,
$\beta_{E\mathbf{k}}$,
and
$\gamma_{E\mathbf{k}}$.
For this system to have a solution, $E$ must satisfy the following
equation:
\begin{equation}
%%%%%%%%%%%%%%%%%%%%%%%%%%%%%%%%%%%%%%%%%%%%%%%%%%
\label{eigenenergy}
%%%%%%%%%%%%%%%%%%%%%%%%%%%%%%%%%%%%%%%%%%%%%%%%%%
E+\epsilon_0 - 2t^2(E+|\Delta|\cos \phi)
\sum \limits_{\mathbf{k}}
\frac {1}{E^2- \epsilon_\mathbf{k}^2-|\Delta|^2}=0.
\end{equation}
This equation may have a subgap solution
$|E_0|<\Delta$:
\begin{eqnarray}
%%%%%%%%%%%%%%%%%%%%%%%%%%%%%%%%%%%%%%%%%%%%%%%%%%
\label{energy_eq}
%%%%%%%%%%%%%%%%%%%%%%%%%%%%%%%%%%%%%%%%%%%%%%%%%%
E_0=-\epsilon_0-\Gamma\frac{E_0+|\Delta|\cos \phi}{\sqrt{\Delta^2-E_0^2}},
\\
\text{ where }
\Gamma=2\pi t^2 \rho_{\rm n} V
\end{eqnarray}
is the width of the impurity level
[$\rho_{\rm n}$
is the density of states at the Fermi energy in metallic state, $V$ is the
sample volume]. There is no more than one subgap solution. It corresponds
to an electron bound to the impurity site. Energy
$E_0$
is zero if
\begin{eqnarray}
\epsilon_0=-\Gamma \cos {\phi}.
\end{eqnarray}
When the order parameter is large ($\Delta \gg \epsilon_0, \Gamma$),
Eq.~(\ref{energy_eq})
may be solved approximately:
\begin{eqnarray}
E_0
\approx
-\epsilon_0-\Gamma \cos {\phi}.
\end{eqnarray}
The energy of the bound state depends on the local value of
$\phi$.
The density of extended states, which will be evaluated below, is also
sensitive to $\phi$. These facts are not surprising: the quantum tunneling
between the impurity and the bands is sensitive to the value of local
charge density, which is proportional to
$\exp(i\phi)$.

\subsection{Density of extended states}

The presence of the impurity not only generates the bound state, but also
affects the density of states in the bulk. For
$|E|>\Delta$
the density of extended states can be calculated with the help of the
following trick
\cite{2-ch}.
All eigenenergies are solutions of
Eq.~(\ref{eigenenergy}).
Thus, the density of states is equal to the ``density of zeros" for the
function:
\begin{eqnarray}
%%%%%%%%%%%%%%%%%%%%%%%%%%%%%%%%%%%%%%%%%%%%%%%%%%
\label{def_F}
%%%%%%%%%%%%%%%%%%%%%%%%%%%%%%%%%%%%%%%%%%%%%%%%%%
F(E)
&=&
-\epsilon_0-E
\\
\nonumber
&+&
2t^2(E+\Delta \cos \phi) \sum \limits_\mathbf{k}
\frac{1}{E^2-\epsilon_\mathbf{k}^2-\Delta^2}.
\end{eqnarray}
However, this function has not only zeros, but poles as well. It is
convenient to define the following polynomial:
\begin{eqnarray}
P(E)=F(E)
\prod \limits_\mathbf{k} (E^2-\epsilon_\mathbf{k}^2-\Delta^2).
\end{eqnarray}
This polynomial has identical set of roots as
$F(E)$,
and no singularities. Thus, the density of states
$\rho(E)$ is equal to
\begin{eqnarray}
%%%%%%%%%%%%%%%%%%%%%%%%%%%%%%%%%%%%%%%%%%%%%%%%%%
\label{density_st}
%%%%%%%%%%%%%%%%%%%%%%%%%%%%%%%%%%%%%%%%%%%%%%%%%%
\rho(E)=\frac 1\pi
\lim_{\omega \rightarrow +0}
	 \frac d{dE} {\rm Im} \left[ \ln P(E+i\omega) \right].
\end{eqnarray}
To prove this formula it is enough to notice that its right-hand side is a
sum of delta-functions
$\sum_n \delta(E-E_n)$,
where
$E_n$
are the roots of
$P(E)$,
and, consequently, of
$F(E)$.

The right-hand side of
Eq.~(\ref{density_st})
can be calculated in the thermodynamic limit. For finite
$\omega>0$
one can replace the sum in the definition of $F$,
Eq.~(\ref{def_F}),
by the integral, which can be evaluated. Finally, we obtain
\begin{eqnarray}
\rho(E)=\rho_0(E)+\frac{1}{\pi}
\frac d{dE} \arctan
\frac{\Gamma (E+\Delta\cos \phi)}{(\epsilon_0+E)\sqrt {E^2-\Delta^2}},
\\
\rho_0 (E) = \rho_{\rm n} \frac{E}{\sqrt{E^2 - \Delta^2}}.
\end{eqnarray}
Here
$\rho_0$
is the usual BCS-like density of states. The density of states
$\rho (E)$
will be used below to calculate the average filling fraction of the
impurity site.

\subsection{Unperturbed many-electron states}

Now, when we have finished describing the single-electron states of the
$U=0$
Hamiltonian, we must construct a set of many-electron states, which will be
the starting point of perturbation theory calculations.

It is assumed that at low temperature all negative-energy extended states
are occupied, and positive-energy extended states are empty. However, the
low-lying subgap states may be empty or occupied, depending on different
conditions (temperature, interaction, magnetic field). To account for these
possibilities we will keep track of four many-electron states
$\left| N_\uparrow, N_\downarrow \right>$,
where numbers
\begin{eqnarray}
N_\sigma = 0 \text{ or } 1
\end{eqnarray}
show how many electrons with spin $\sigma$ sit at the impurity bound state.
The energy of
$\left| N_\uparrow, N_\downarrow \right>$
equals to:
\begin{eqnarray}
%%%%%%%%%%%%%%%%%%%%%%%%%%%%%%%%%%%%%%%%%%%%%%%%%%
\label{E_bare}
%%%%%%%%%%%%%%%%%%%%%%%%%%%%%%%%%%%%%%%%%%%%%%%%%%
E = \sum_\sigma E_\sigma,
\\
%%%%%%%%%%%%%%%%%%%%%%%%%%%%%%%%%%%%%%%%%%%%%%%%%%
\label{E_sigma_bare}
%%%%%%%%%%%%%%%%%%%%%%%%%%%%%%%%%%%%%%%%%%%%%%%%%%
E_\sigma
=
E_0 N_\sigma
+
\int {\bar E} \rho ({\bar E}) \Theta(-\Delta -{\bar E}) d{\bar E}.
\end{eqnarray}
Here
$\Theta(x)$
is the step-function.

\subsection{Filling fraction of the impurity site}
%%%%%%%%%%%%%%%%%%%%%%%%%%%%%%%%%%%%%%%%%%%%%%%%%%
\label{fillin_frac}
%%%%%%%%%%%%%%%%%%%%%%%%%%%%%%%%%%%%%%%%%%%%%%%%%%

To apply the perturbation theory, we will need the following matrix
element:
\begin{eqnarray}
\langle n_\sigma \rangle
=
\left< N_\uparrow, N_\downarrow \right|
	d_\sigma^{\dagger}d_\sigma^{\vphantom{\dagger}}
\left| N_\uparrow, N_\downarrow \right>.
\end{eqnarray}
Since the non-interacting Hamiltonian does not couple different spin
projections,
$\langle n_\sigma \rangle$
depends on
$N_\sigma$,
but not on
$N_{-\sigma}$.
Physically,
$\langle n_\sigma \rangle$
is equal to probability of finding electron with spin $\sigma$ on the
impurity. Note that this probability is not equal to
$N_\sigma$.

To calculate
$\langle n_\sigma \rangle$
it is convenient to use the Hellmann-Feynman theorem, which states that, if
$\left| \psi_n (\lambda) \right>$
is an eigenstate of a Hamiltonian
$H=H(\lambda)$:
\begin{eqnarray}
H (\lambda) \left| \psi_n (\lambda) \right>
=
E_n (\lambda) \left| \psi_n (\lambda) \right>,
\end{eqnarray}
where $\lambda$ is some parameter, then
\begin{eqnarray}
\frac{\partial E_n}{\partial \lambda}
=
\left< \psi_n \right|
	\frac{\partial H}{\partial \lambda}
\left| \psi_n \right>.
\end{eqnarray}
Using the latter equation, we can write
\begin{eqnarray}
%%%%%%%%%%%%%%%%%%%%%%%%%%%%%%%%%%%%%%%%%%%%%%%%%%
\label{hf}
%%%%%%%%%%%%%%%%%%%%%%%%%%%%%%%%%%%%%%%%%%%%%%%%%%
\langle n_\sigma \rangle
=
-\frac {\partial E_\sigma}{\partial \epsilon_0},
\end{eqnarray}
where
$E_\sigma$
is determined by
Eq.~(\ref{E_sigma_bare}).
The formula for
$\langle n_\sigma \rangle$
has two terms:
\begin{eqnarray}
%%%%%%%%%%%%%%%%%%%%%%%%%%%%%%%%%%%%%%%%%%%%%%%%%%
\label{filling_fraction}
%%%%%%%%%%%%%%%%%%%%%%%%%%%%%%%%%%%%%%%%%%%%%%%%%%
\langle n_\sigma \rangle
=
n_0+n_1 N_\sigma,
\\
%%%%%%%%%%%%%%%%%%%%%%%%%%%%%%%%%%%%%%%%%%%%%%%%%%
\label{filling_fraction_n0}
%%%%%%%%%%%%%%%%%%%%%%%%%%%%%%%%%%%%%%%%%%%%%%%%%%
n_0
=
\int {\bar E}
\frac{
	\partial \rho ({\bar E})
     }
     {
	\partial \epsilon_0
     }
\Theta(- \Delta -{\bar E})
d{\bar E},
\\
n_1
=
\frac{\partial E_0}{\partial \epsilon_0}.
\end{eqnarray}
The quantity
$n_0$,
Eq.~(\ref{filling_fraction_n0})
is independent of
$N_\sigma$,
and identical for both spin projections. It is always non-zero as long
as the impurity level has finite hybridization with the band electrons.
General analytic calculations for
$n_0$
are quite cumbersome, however, some simple equations can be obtained in the
limit
$\epsilon_0\ll \Delta$.
When
$\phi=\frac {\pi}{2}$,
for arbitrary $\Gamma$ the following relation can be derived:
\begin{eqnarray}
n_0(\phi=\frac {\pi}{2})=\frac {1}{2} \frac {\Gamma}{\Gamma+\Delta}.
\end{eqnarray}
If we assume further that
$\Gamma < \Delta$,
then
\begin{eqnarray}
n_0(\phi=0)\simeq \frac{\Gamma}{\Delta}(\frac {1}{2}- \frac {1}{\pi}).
%\\
%n_0(\phi=\pi)\simeq\frac {1}{2} \frac {\Gamma \Delta }{\Gamma^2+\Delta^2}
\end{eqnarray}

The second term,
$N_\sigma n_1$,
obviously depends on the occupation of the bound state level by an electron
with spin $\sigma$. Using
Eq.~(\ref{energy_eq})
we derive
\begin{eqnarray}
n_1
=
\frac{
	(\Delta^2-E_0^2)^\frac 32
     }
     {
	(\Delta^2-E_0^2)^\frac 32+\Gamma \Delta (\Delta+E_0\cos \phi)
     }.
\end{eqnarray}
%%%%%%%%%%%%%%%%%%%%%%%%%%%%%%%%%%%%%%%%%%%%%%%%%%
\begin{figure}[t!]
\center
\includegraphics [width=8.5cm, height=4.5cm]{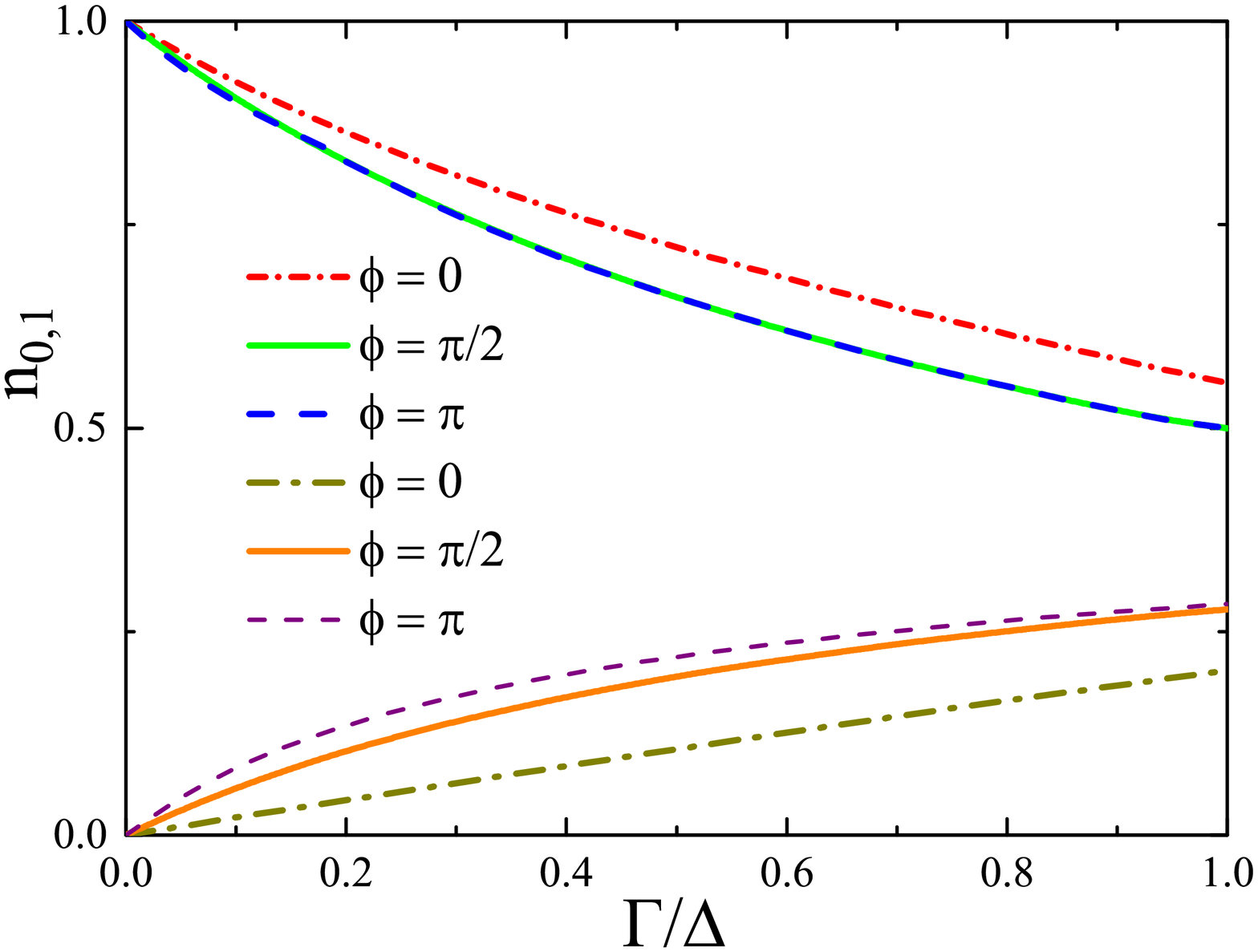}
\caption{(Color online)
The dependence of
$n_{0}$
and
$n_{1}$
versus the impurity level width $\Gamma$ for different values of $\phi$
(phase $\phi$ characterizes the impurity location
${\bf R}_{\rm imp}$
relative to the charge-density wave). For calculations
$\epsilon_0=0.2\Gamma$
was chosen. All curves for
$n_{1}$
start from 1
at
$\Gamma = 0$,
and decrease when $\Gamma$ grows. The curves for
$n_0$
start from zero at
$\Gamma = 0$,
and grow when $\Gamma$ grows.
}
%%%%%%%%%%%%%%%%%%%%%%%%%%%%%%%%%%%%%%%%%%%%%%%%%%
\label{stm_exp}
%%%%%%%%%%%%%%%%%%%%%%%%%%%%%%%%%%%%%%%%%%%%%%%%%%
\end{figure}
%%%%%%%%%%%%%%%%%%%%%%%%%%%%%%%%%%%%%%%%%%%%%%%%%%
%

Both
$n_0$
and
$n_1$
are plotted in
Fig.~\ref{stm_exp}.
Examining this figure one can notice that, if
$E_0<0$
and
$\epsilon_0,\Gamma < \Delta$,
then the contribution of zone electrons
$n_0$
is much smaller than contribution of the localized electrons
$n_1$.

\subsection{Perturbation theory}

In this subsection we calculate first-order correction to the energy of
$\left| N_\uparrow, N_\downarrow \right>$
induced by small, but finite $U$. It is given by the following matrix
element:
\begin{eqnarray}
\Delta E
=
U
\left< N_\uparrow, N_\downarrow \right|
	d^{\dagger}_{\uparrow}d_{\uparrow}^{\vphantom{\dagger}}
	d^{\dagger}_{\downarrow}d_{\downarrow}^{\vphantom{\dagger}}
\left| N_\uparrow, N_\downarrow \right>
=
U \langle n_\uparrow \rangle \langle n_\downarrow \rangle.
\end{eqnarray}
Since the non-perturbed Hamiltonian does not couple spin projections, the
latter matrix element factorizes into a product
$\langle n_\uparrow \rangle \langle n_\downarrow \rangle$,
which can be evaluated easily:
\begin{eqnarray}
%%%%%%%%%%%%%%%%%%%%%%%%%%%%%%%%%%%%%%%%%%%%%%%%%%
\label{N=0}
%%%%%%%%%%%%%%%%%%%%%%%%%%%%%%%%%%%%%%%%%%%%%%%%%%
\Delta_0 E=Un_0^2
&\text{ for }& N = 0, S_z = 0,
\\
%%%%%%%%%%%%%%%%%%%%%%%%%%%%%%%%%%%%%%%%%%%%%%%%%%
\label{N=1}
%%%%%%%%%%%%%%%%%%%%%%%%%%%%%%%%%%%%%%%%%%%%%%%%%%
\Delta_1 E=Un_1n_0+Un_0^2
&\text{ for }& N = 1, S_z = \pm 1/2,\quad
\\
%%%%%%%%%%%%%%%%%%%%%%%%%%%%%%%%%%%%%%%%%%%%%%%%%%
\label{N=2}
%%%%%%%%%%%%%%%%%%%%%%%%%%%%%%%%%%%%%%%%%%%%%%%%%%
\Delta_2 E=U(n_1+n_0)^2
&\text{ for }& N = 2, S_z = 0,
\end{eqnarray}
where
$\Delta_{N} E$
is the correction for the state with $N$ electrons on the bound state. Of
the three possibilities presented by
Eqs.~(\ref{N=0}),
(\ref{N=1}),
and
(\ref{N=2}),
only
$N=1$
case corresponds to a magnetic state with non-zero spin. This state becomes
the ground state if its energy
$E_0+\Delta_1 E$
is lower than both the energy
$2E_0+\Delta_2 E$
of the state with two electrons bound to the impurity, and the energy
$\Delta_0 E$
for the state with zero electrons at the bound state. Therefore, the ground
state is magnetic if
\begin{eqnarray}\label{pm_cond}
-Un_1^2-Un_0n_1<E_0<-Un_0n_1.
\end{eqnarray}
For large
$\Delta \gg \epsilon_0, \Gamma$
the latter condition is equivalent to:
\begin{eqnarray}
%%%%%%%%%%%%%%%%%%%%%%%%%%%%%%%%%%%%%%%%%%%%%%%%%%
\label{pm_cond_approx}
%%%%%%%%%%%%%%%%%%%%%%%%%%%%%%%%%%%%%%%%%%%%%%%%%%
U>\epsilon_0+\Gamma \cos {\phi}>0.
\end{eqnarray}
Equation~(\ref{pm_cond})
allows us to map numerically the phase diagram of the impurity.
\begin{figure}[t!]
\center
\includegraphics [width=8.5cm, height=4.5cm]{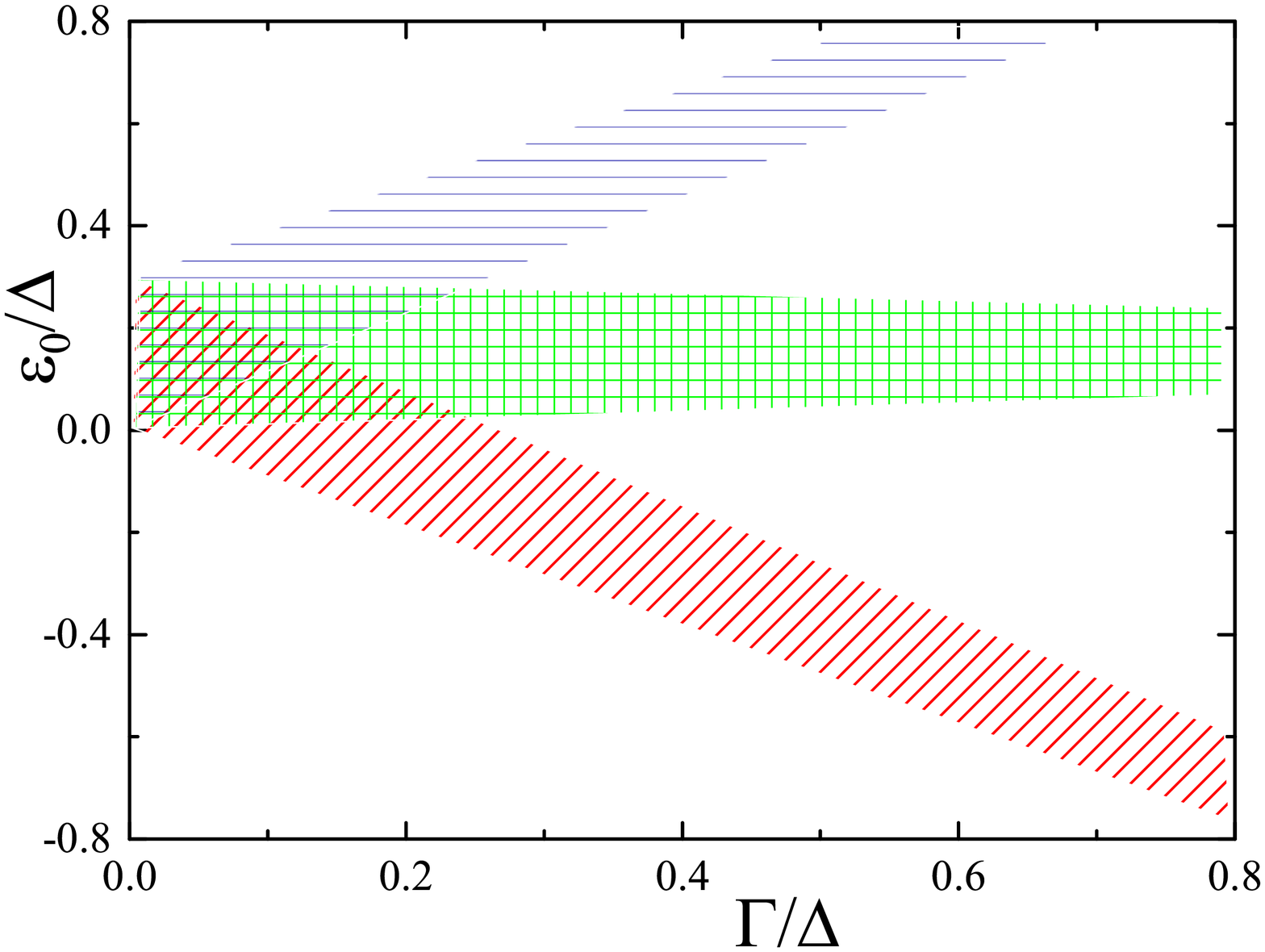}
\caption{(Color online) The phase diagram of a single impurity on the plane
$(\Gamma/\Delta, \epsilon_0/\Delta)$.
The value of $U$ is equal to
$0.3\Delta$
The hatched areas correspond to the magnetic phase. The difference between
the three hatched areas is the value of phase $\phi$, which depends on the
location of the impurity relative to the density wave.
Horizontally hatched area (blue) represents the magnetic phase for
$\phi = 0$.
Crisscrossed area (green) represents the magnetic phase for
$\phi = \pi /2$.
The area hatched by slanted (red) lines represents the magnetic phase for
$\phi = \pi$.
}
%%%%%%%%%%%%%%%%%%%%%%%%%%%%%%%%%%%%%%%%%%%%%%%%%%
\label{phase_diag}
%%%%%%%%%%%%%%%%%%%%%%%%%%%%%%%%%%%%%%%%%%%%%%%%%%
\end{figure}

The phase diagrams on the plane
$(\Gamma/\Delta, \epsilon_0/\Delta)$
for different $\phi$ are presented in
Fig.~\ref{phase_diag}.
The value of $\phi$ affects the details of the phase diagram, however, the
magnetic phase exist for any $\phi$.

This phase diagram is valid as long as the perturbation theory is
justified. One can use the perturbation theory if the bound state energy
$E_0$
lies sufficiently far from the edges of the continuous spectrum. Thus, for
small $U$ our phase diagram is valid even for large
$\epsilon_0, \Gamma \lesssim \Delta$.
\begin{figure}[t!]
\center
\includegraphics [width=8.5cm, height=4.5cm]{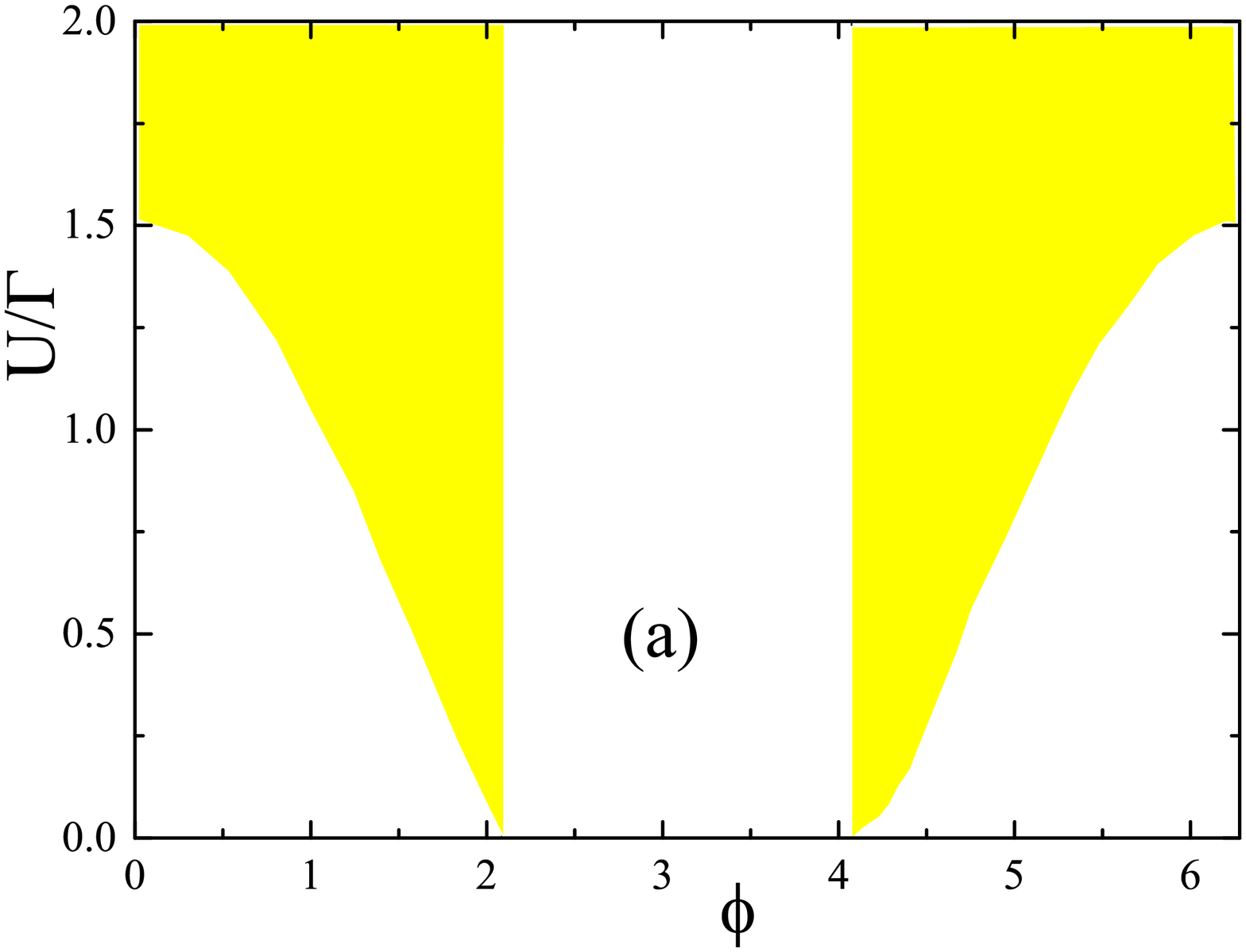}
\includegraphics [width=8.5cm, height=4.5cm]{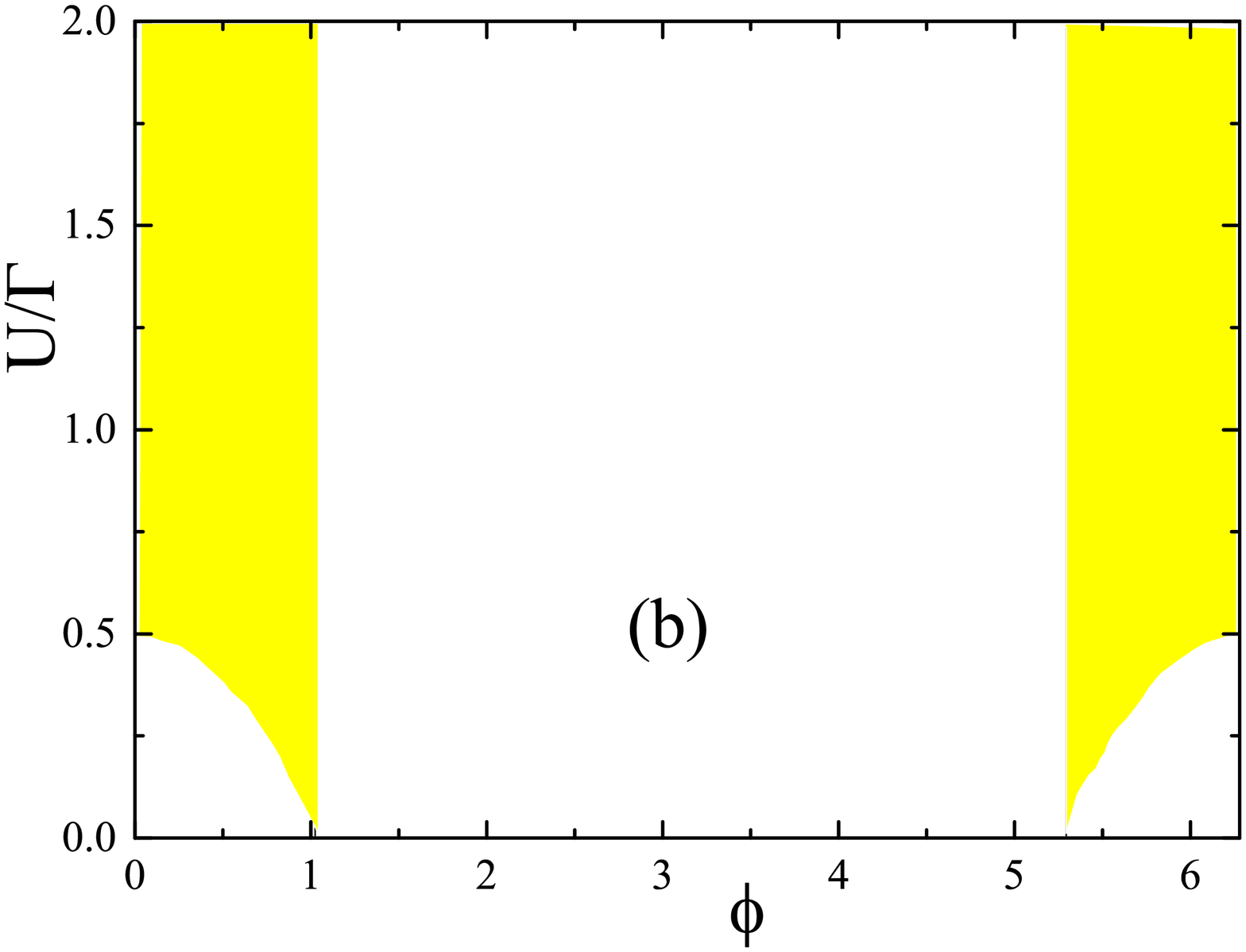}
\includegraphics [width=8.5cm, height=4.5cm]{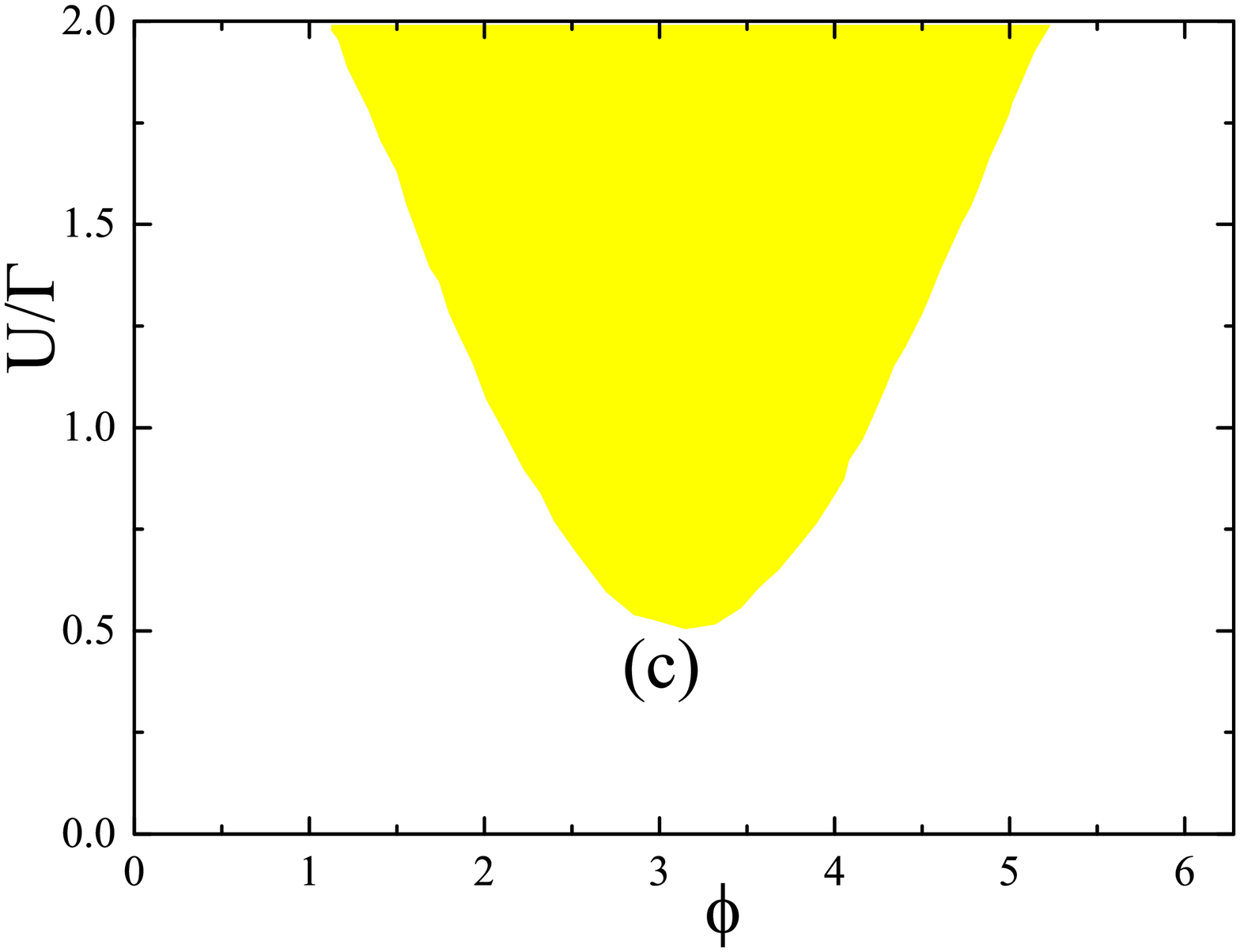}
\caption{(Color online) The phase diagrams of the impurity for three
different value of the on-site potential $\epsilon_0$. Vertical axis is the
Coulomb repulsion $U$, the horizontal axis is $\phi$. Colored (yellow) area
shows the magnetic phase.
Panel (a) corresponds to
$\epsilon_0=-0.5\Gamma$,
panel (b) corresponds to
$\epsilon_0=0.5\Gamma$,
panel (c) corresponds to
$\epsilon_0=1.5\Gamma$.
}
\label{U_phi}
\end{figure}

In
Fig.~\ref{U_phi}
the phase diagram of a single impurity is shown on a different plane. This
time, the horizontal axis represents $\phi$, the vertical axis --
interaction parameter
$U/\Gamma$.
The magnetic phase corresponds to the colored (yellow) area. It is easy to
see that, if
$|\epsilon_0|<\Gamma$,
then for any non-zero value of $U$ there is finite interval of $\phi$,
where the impurity is magnetic. For relatively small interaction
$U \ll \sqrt{\Gamma^2-\epsilon_0^2}$
the width of such an interval is determined by the following formula
\begin{equation}
\delta \phi_U=\frac{2U}{\sqrt{\Gamma^2-\epsilon_0^2}}.
\end{equation}
If the impurities are randomly distributed along the CDW then even weak
repulsion can magnetize at least part of them. In the next section we will
discuss how this affects the experimentally observed quantities.

\section{Thermodynamic properties of an impurity ensemble}
%%%%%%%%%%%%%%%%%%%%%%%%%%%%%%%%%%%%%%%%%%%%%%%%%%
\label{sect::ensamble}
%%%%%%%%%%%%%%%%%%%%%%%%%%%%%%%%%%%%%%%%%%%%%%%%%%

In the previous section we learned that a weakly interacting impurity in a
CDW material may have magnetic ground state. In this section we will
demonstrate that in a system with finite concentration of such impurities
the magnetic susceptibility diverges at zero temperature, while the heat
capacity at finite temperature demonstrates pronounced peak at weak
magnetic field.

\subsection{Susceptibility of a single impurity}

We start our analysis with calculation of the partition function of a
single impurity placed in magnetic field. Only
$\left| N_\uparrow, N_\downarrow \right>$
states contribute to the thermodynamic properties at low temperature and
magnetic field. Their energies are:
\begin{eqnarray}
%%%%%%%%%%%%%%%%%%%%%%%%%%%%%%%%%%%%%%%%%%%%%%%%%%
\label{imp_energies1}
%%%%%%%%%%%%%%%%%%%%%%%%%%%%%%%%%%%%%%%%%%%%%%%%%%
\epsilon_1=Un_0^2, \\
%%%%%%%%%%%%%%%%%%%%%%%%%%%%%%%%%%%%%%%%%%%%%%%%%%
\label{imp_energies2}
%%%%%%%%%%%%%%%%%%%%%%%%%%%%%%%%%%%%%%%%%%%%%%%%%%
\epsilon_{2,3}=E_0+Un_0(n_1+n_0)\pm h,\\
%%%%%%%%%%%%%%%%%%%%%%%%%%%%%%%%%%%%%%%%%%%%%%%%%%
\label{imp_energies3}
%%%%%%%%%%%%%%%%%%%%%%%%%%%%%%%%%%%%%%%%%%%%%%%%%%
\epsilon_4=2E_0+U(n_1+n_0)^2,
\end{eqnarray}
where
$\epsilon_1$
is the energy of the state with zero localized electrons on the impurity.
This energy is determined by Coulomb repulsion $Un_0^2$ between bulk
electrons tunneled to the impurity site. Energies
$\epsilon_{2,3}$
correspond to the states with one localized electron. They have finite
Zeeman energy
$h_{2,3}=\pm\mu_{\rm B} B$.
When there are two electrons on the impurity site, their energy
$\epsilon_4$
is composed of two contributions: the single-electron energy
$2 E_0$,
and the electron-electron interaction energy
$U(n_0 + n_1)^2$.
In the above expressions we did not include explicitly the kinetic energy
of the zone electrons. Since the occupation numbers of bulk states does not
change at low $T$ and $h$, this portion of energy is identical for all four
states
$\left| N_\uparrow, N_\downarrow \right>$.

The corresponding partition function is:
\begin{eqnarray}
%%%%%%%%%%%%%%%%%%%%%%%%%%%%%%%%%%%%%%%%%%%%%%%%%%
\label{Z}
%%%%%%%%%%%%%%%%%%%%%%%%%%%%%%%%%%%%%%%%%%%%%%%%%%
Z=\sum \limits_{i=1}^4  e^{-\beta \epsilon_i},
\text{ where }
\beta = 1/T.
\end{eqnarray}
It can be used to find the free energy
$F=-T\ln {Z}$,
which, in turn, is used to calculate the magnetic susceptibility of the
impurity:
\begin{eqnarray}
%%%%%%%%%%%%%%%%%%%%%%%%%%%%%%%%%%%%%%%%%%%%%%%%%%
\label{chi_general}
%%%%%%%%%%%%%%%%%%%%%%%%%%%%%%%%%%%%%%%%%%%%%%%%%%
\chi=-\frac{\partial^2 F}{\partial h^2}|_{h=0}
=\frac {2 \beta}{e^{\beta {\overline E}_0}
+
2+e^{-\beta({\overline E}_0+{\overline U})}}.
\end{eqnarray}
Within the framework of our formalism the renormalized energies in this
equation equal to:
\begin{eqnarray}
%%%%%%%%%%%%%%%%%%%%%%%%%%%%%%%%%%%%%%%%%%%%%%%%%%
\label{eff_E0}
%%%%%%%%%%%%%%%%%%%%%%%%%%%%%%%%%%%%%%%%%%%%%%%%%%
{\overline E}_0 = E_0+Un_0n_1,
\\
%%%%%%%%%%%%%%%%%%%%%%%%%%%%%%%%%%%%%%%%%%%%%%%%%%
\label{eff_U}
%%%%%%%%%%%%%%%%%%%%%%%%%%%%%%%%%%%%%%%%%%%%%%%%%%
{\overline U} = Un_1^2
\end{eqnarray}
These expressions are obtained using the perturbation theory.
However,
Eq.~(\ref{chi_general})
is more general: it captures the physics of an Anderson impurity in an
insulating environment. This equation retains its physical meaning even
when the perturbation theory is invalid. In this case, of course, simple
perturbation theory results,
Eqs.~(\ref{eff_E0})
and
(\ref{eff_U}),
must be discarded. Instead, parameters
${\overline E}_0$
and
${\overline U}$
should be derived using a more advanced technique (e.g., mean field
theory).

At small $T$ the susceptibility diverges:
\begin{eqnarray}
\chi(T) \approx \frac{1}{T},
\end{eqnarray}
provided that
${\overline E}_0<0$
and
${\overline  E}_0+ {\overline U}>0$
at the same time. Otherwise, the susceptibility vanishes at
$T=0$.

\subsection{Susceptibility of the impurity ensemble}
%%%%%%%%%%%%%%%%%%%%%%%%%%%%%%%%%%%%%%%%%%%%%%%%%%
\label{tls::susceptibility_ensemble}
%%%%%%%%%%%%%%%%%%%%%%%%%%%%%%%%%%%%%%%%%%%%%%%%%%

Of course, in a sample numerous impurities exist. If the impurities are not
too dense, they can be treated independently. To study a macroscopic sample
in the diluted limit single-impurity properties must be averaged over an
ensemble of impurities.

To perform the averaging we assume that all impurities have the same values
of the ``internal" parameters $U$, $\Gamma$, and
$\epsilon_0$.
The phase $\phi$, on the other hand, will be treated as a random variable
homogeneously distributed over the interval of
$(0, 2\pi)$.
This means that the locations of the impurities do not correlate with the
charge-density modulations in the CDW phase, which is equivalent to
weak pinning of the CDW by the impurities. (The case of strong pinning
will be briefly discussed in
Sec.~\ref{sect::discussion}.)

\begin{figure}[t!]
\center
\includegraphics [width=8.5cm, height=4.5cm]{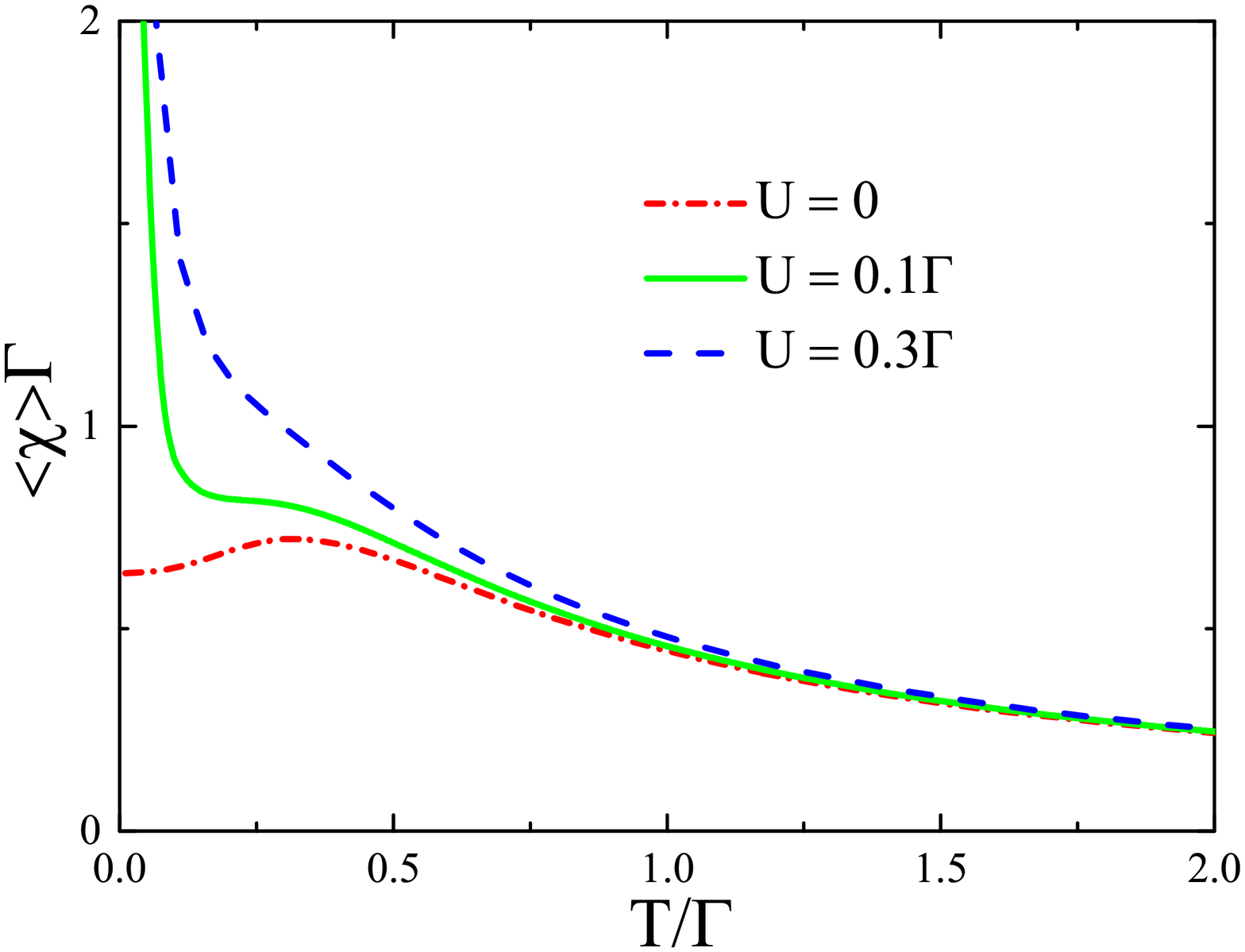}
\caption{(Color online) The ensemble-averaged magnetic susceptibility
$\langle \chi \rangle$
versus temperature $T$ for different values of the Coulomb repulsion
$U/\Gamma$.
Other parameters are:
$\epsilon_0=0.1\Gamma$,
Note that
$\Gamma > \epsilon_0$.
When this condition is met, the susceptibility diverges at small $T$ for
arbitrary weak $U$. By contrast, the ensemble of
$U=0$
impurities has finite susceptibility at
$T=0$.
}
%%%%%%%%%%%%%%%%%%%%%%%%%%%%%%%%%%%%%%%%%%%%%%%%%%
\label{xi}
%%%%%%%%%%%%%%%%%%%%%%%%%%%%%%%%%%%%%%%%%%%%%%%%%%
\end{figure}

The ensemble of this type has an interesting property: if
$\Gamma>\epsilon_0$,
then for any non-zero $U$ a finite fraction of the impurities are in the
magnetic state. In other words, arbitrary small interaction is sufficient
to generate the divergent susceptibility of the ensemble. The results of
numerical calculations illustrating this point are presented in
Fig.~\ref{xi}.
There the dependence of the ensemble-averaged magnetic susceptibility
$\langle \chi \rangle$
is shown as a function of temperature for different values of the repulsion
$U$. If there is no interaction, then
$\langle \chi(T) \rangle$
is finite for any temperature. However, the susceptibility demonstrates
$1/T$
divergence (the Curie's law) even for weak $U$. The smaller the
interaction, the lower the temperature at which the susceptibility starts
to diverge.

\subsection{Ensemble heat capacity}
%%%%%%%%%%%%%%%%%%%%%%%%%%%%%%%%%%%%%%%%%%%%%%%%%%
\label{heat_cap}
%%%%%%%%%%%%%%%%%%%%%%%%%%%%%%%%%%%%%%%%%%%%%%%%%%

\begin{figure}[t!]
\center
\includegraphics [width=8.5cm, height=4.5cm]{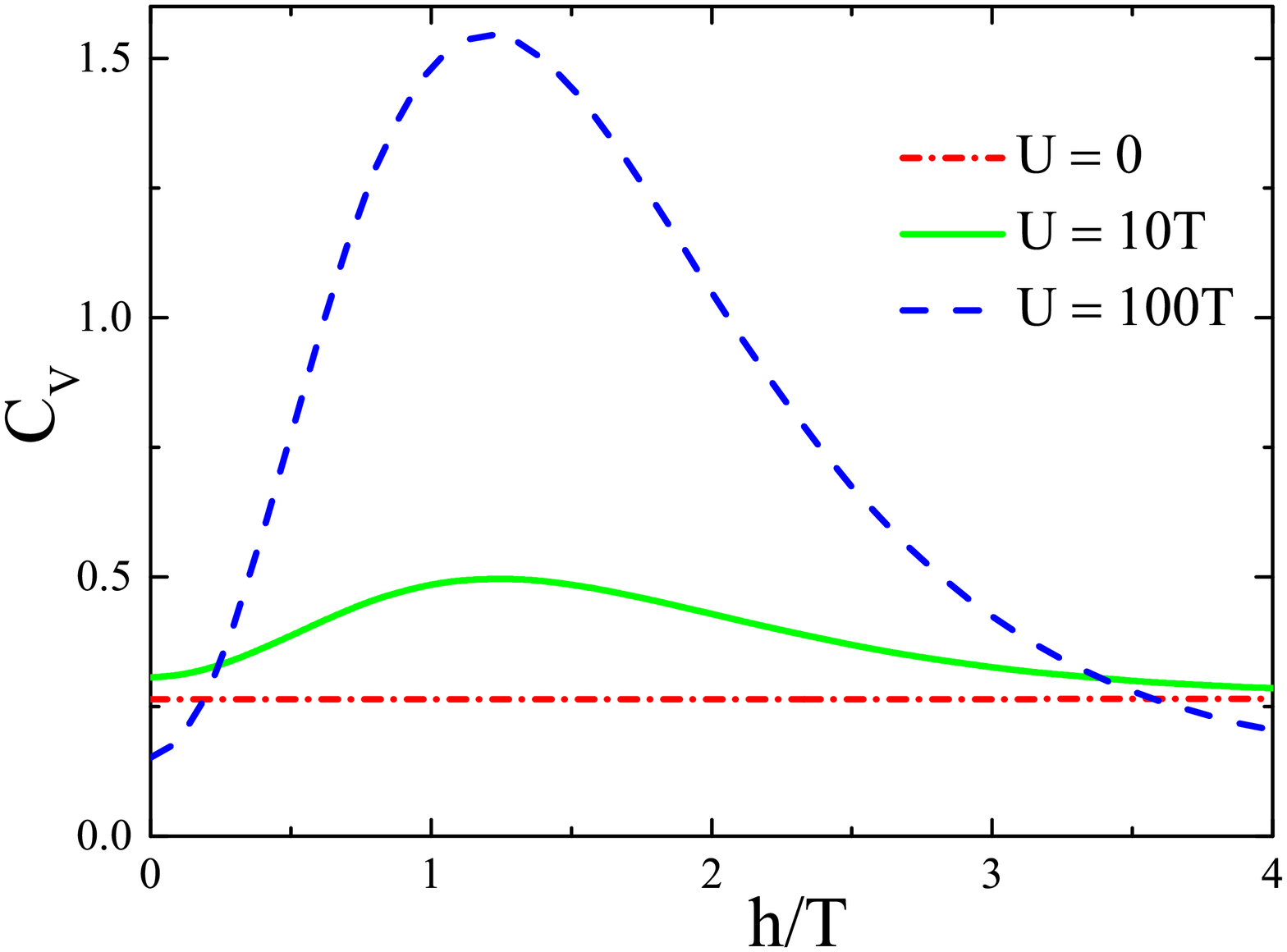}
\caption{(Color online) The heat capacity
$C_{\rm V}$
of the impurity ensemble versus Zeeman energy $h$ for different values of
the interaction parameter $U$. For calculations we used
$\Gamma=50T$
and
$\epsilon_0=T$.
If
$U=0$,
the heat capacity is virtually field independent at small $h$. For
non-zero $U$ finite concentration of the magnetized impurities emerge, and
$C_{\rm V}$
develops maximum at weak field
$h \sim T$.
We see also that the interaction $U$ significantly enhances the weak field
heat capacity.
%%%%%%%%%%%%%%%%%%%%%%%%%%%%%%%%%%%%%%%%%%%%%%%%%%
\label{Cv_vs_h}
%%%%%%%%%%%%%%%%%%%%%%%%%%%%%%%%%%%%%%%%%%%%%%%%%%
}
\end{figure}

Besides the susceptibility, the heat capacity of o-TaS$_3$ as a function of
temperature and magnetic field has been
measured~\cite{biljakovic_otas3_epl,cdw_heat_capacity}.
It is not difficult to extend the formalism of the previous subsection for
calculation of the impurity ensemble heat capacity
$C_{\rm V}$.
It equals to:
\begin{eqnarray}
C_{\rm V} (T, h)
=
-T\frac {\partial^2 \langle F  (T, h)\rangle}{{\partial T}^2},
\end{eqnarray}
where
$\langle F \rangle$
denotes the ensemble-averaged free energy.

The dependence of
$C_{\rm V}$
on Zeeman energy $h$ for fixed temperature is presented in
Fig.~\ref{Cv_vs_h}.
As we can see, if
$U=0$,
the heat capacity is quite insensitive to weak magnetic field. However, for
non-zero interaction
$C_{\rm V}$
becomes a non-monotonous function of the magnetic field, with maximum near
$h \sim T$.
The maximum is associated with the contributions of those impurities on
which exactly one electron resides. Since the number of such impurities in
the ensemble grows when $U$ increases, the weak field heat capacity may be
significantly enhanced by the interaction, as one can see from the graphs
in
Fig.~\ref{Cv_vs_h}

\section{Discussion}
%%%%%%%%%%%%%%%%%%%%%%%%%%%%%%%%%%%%%%%%%%%%%%%%%%
\label{sect::discussion}
%%%%%%%%%%%%%%%%%%%%%%%%%%%%%%%%%%%%%%%%%%%%%%%%%%

\subsection{Comparison to experiment}

Above we proposed a mechanism describing the emergence of localized
magnetic momenta in a material with CDW order. Within the framework of our
model we were able to show that the magnetic susceptibility of the system
diverges at low $T$. This behavior resembles the phenomenology observed in
experiments on o-TaS$_3$,
Ref.~\onlinecite{biljakovic_otas3_epl},
and YTe$_3$ and LaTe$_3$,
Ref.~\onlinecite{nancy_ru_thesis,nancy_ru_prb}.
Specifically, the susceptibility of
o-TaS$_3$
is virtually temperature-independent for a broad range of temperatures,
both above and below its CDW transition temperature of
$T_{\rm CDW} = 218$\,K.
However, below
$\sim$60\,K
the susceptibility diverges as
$T\rightarrow 0$.
Ref.~\onlinecite{biljakovic_otas3_epl}
explored the details of this divergence.

Non-magnetic tritellurides
YTe$_3$ and LaTe$_3$
demonstrated diamagnetic susceptibility with weak temperature
dependence~\cite{nancy_ru_prb,nancy_ru_thesis}.
However, below
$\sim 10$\,K
divergent paramagnetic contribution appears, see Fig.~6.1 of
Ref.~\onlinecite{nancy_ru_thesis}.
While that contribution has been dismissed as being due to contamination by
magnetic atoms, no experimental proof to this statement has been offered.
Obvious similarity between the behavior of the quasi-two-dimensional and
quasi-one-dimensional materials suggests that, beside contamination, other
options must not be dismissed off hand. Clearly, the susceptibility
divergence of both
TaS$_3$
and the tritellurides is in agreement with the conclusions of our study.

Another interesting experimental feature of o-TaS$_3$ is the sensitivity of
its heat capacity to low magnetic field. In
Ref.~\onlinecite{biljakovic_otas3_others}
it has been reported that at
$T=0.1$\,K
the field-sensitive contribution to
$C_{\rm V}$
passes through maximum when the magnetic field $B$ is equal to
0.1\,T.
Such a magnetic field corresponds to the Zeeman energy
$h=\mu_B B=0.065$\,K.
This is consistent with the heat capacity behavior presented in our
Fig.~\ref{Cv_vs_h}:
if
$U > 0$,
then
$C_{\rm V}(h)$
has a pronounced maximum at
$h \sim T$.

\subsection{Spin-spin interaction}

Our model qualitatively reproduces both the susceptibility divergence and
the sensitivity of the heat capacity to the magnetic field, yet, there are
obvious discrepancies with the data. On experiment the susceptibility
diverged with fractional exponent; the heat capacity demonstrated
hysteresis when magnetic field was
varied.~\cite{biljakovic_otas3_others,biljakovic_otas3_epl}.
To describe these phenomena the model of non-interacting impurities is
insufficient. It is likely that interaction between impurity spins must be
accounted for. For example, random-exchange antiferromagnetic Heisenberg
model has been mentioned in
Ref.~\onlinecite{biljakovic_otas3_epl}
as a possible low-temperature effective theory responsible for the
fractional exponent in the susceptibility data. However, a study of such
an interaction is beyond the scope of this paper.

\subsection{Strength of the effective Coulomb interaction}
%%%%%%%%%%%%%%%%%%%%%%%%%%%%%%%%%%%%%%%%%%%%%%%%%%
\label{tls::strength_U}
%%%%%%%%%%%%%%%%%%%%%%%%%%%%%%%%%%%%%%%%%%%%%%%%%%

Our investigation demonstrates that the repulsion between electrons on
the impurity site $U$ is of crucial importance for non-trivial magnetic
properties of the CDW. The non-zero $U$ is necessary to explain both the
divergence of the magnetic susceptibility (see
Fig.~\ref{xi}),
and the low field peak of the heat capacity (see
Fig.~\ref{Cv_vs_h}).

Given such a sensitivity to $U$, we would like to discuss its value.
Clearly, we can give reasonable estimate for the interaction strength only
when the nature of the impurities is known. If we deal with an atomic
impurity, the value of $U$ may be as high as eV or several eV due to strong
localization of the atomic orbitals. For such a high value of the
interaction the perturbation theory is not applicable. Instead, one should
use the mean field theory to find the impurity magnetization
self-consistently. As for the main conclusions of our study, they remain
unchanged. However, if such high-$U$ impurities were indeed present in the
material, they might form localized magnetic momenta even in the metallic
phase (provided that
$\epsilon_0 < 0$)
even in the metallic phase. Yet, it appears that at high temperature
nothing anomalous has been reported.

On the other hand, visually examining experimental data in Fig.~1a of
Ref.~\onlinecite{biljakovic_otas3_epl},
we notice that the susceptibility of
TaS$_3$
starts to grow below
$\sim 100$\,K.
If we interpret this observation within the framework of our model, the
following rough estimate
$U \sim 100\,{\rm K} = 10\,{\rm meV}$
can be made. Such a small value of $U$ could mean one of two things. Either
(i) the effective value of $U$ experiences strong renormalization due to,
for example, hopping to the bulk, or (ii) that ``the impurities" are, in
fact, shallow defects due to structural imperfections.

The decreasing renormalization of $U$ due to hopping is quite expected.
However, to justify case (i) the renormalization must be very strong: about
two orders of magnitude, from
$\sim$\,eV
to
$\sim$\,10\,meV.
It is not clear if this could be rationalized for a real material. However,
if it is indeed possible, then our formalism may be straightforwardly
applied to such a system.

Regarding case (ii), our model cannot be immediately applied to such
``impurities": the hybridization Hamiltonian,
Eq.~(\ref{H_hop}),
corresponds to a very localized impurity state, which hybridizes with the
band states at
${\bf R} = {\bf R}_{\rm imp}$
only. However, the main conclusions of our analysis endure.

To prove the latter statement, consider the following reasoning. Let us
model the crystal imperfections by spatial variation of the
single-particle potential
$V = V({\bf R})$.
It is finite near the imperfection:
\begin{eqnarray}
V({\bf R}) \sim V_0 < 0,
\text{ if }
|{\bf R}| < R_0,
\end{eqnarray}
and zero otherwise. In a gapful environment, like CDW material, if
$V_0$
is sufficiently deep, and
$R_0$
is sufficiently large, the imperfection may host a subgap localized state.
Due to very extended wave function the effective $U$ for such a state is
low. However, for a shallow subgap state even weak $U$ might be sufficient
to push the second electron out, generating a spinful ground state. An
ensemble of these states would demonstrate both the divergent
susceptibility, and the low field peak of
$C_{\rm V} (h)$.

\subsection{Strong versus weak pinning}

When we averaged over the impurity ensemble in
subsection~\ref{tls::susceptibility_ensemble},
we assumed that the positions of impurities and the order parameter phase
$\phi$ are not correlated. This assumption is equivalent to homogeneous
distribution of $\phi$ in the impurity ensemble.

However, in general, the CDW pinning introduces the correlation between the
impurity position and the phase $\phi$. Indeed, an impurity and the CDW
interact. The corresponding pinning energy is a periodic function of
$\phi$. An impurity ensemble introduces local distortions to the CDW to
minimize the pinning energy at the expense of the CDW elastic energy.
For weakly pinned CDW this distortion is weak. In this situation the
correlation between impurity positions and $\phi$ may be neglected.

In the opposite limit of strong pinning a given impurity strongly
distorts the CDW to choose a particular value of $\phi$, which minimizes
the pinning energy function. In such a regime the assumption of homogeneous
distribution of $\phi$ is, clearly, invalid. Instead, a distribution
function would concentrate around a particular value (or values) of $\phi$.
This, however, does not affect the major conclusions of our study.
Specifically, both the divergence of the susceptibility and the low field
maximum of the heat capacity are consequences of the impurities hosting
spins. To generate at least some amount of the spinful impurities the
ensemble we have introduced in
subsection~\ref{tls::susceptibility_ensemble}
requires arbitrary weak $U$. For different distribution function it may be
necessary for $U$ to exceed some critical strength
$U_c$.
If
$U > U_c$,
then both divergent susceptibility and non-monotonous field-dependent heat
capacity should be expected.

\subsection{Mechanism of Vakhitov et al.}

A possible theoretical mechanism explaining generation of the magnetic
moments in CDW state has been proposed in
Ref.~\onlinecite{vakhitov}
by Vakhitov and co-authors. In that reference a strongly anisotropic
quasi-one-dimensional (Q1D) metal interacting with phonon mode has been
studied. The metal undergoes the Peierls transition at some finite
temperature. It was demonstrated that, under suitable conditions, an
impurity introduced into such a system traps an unpaired electron. These
impurities, randomly scattered over the sample, are responsible for the
low-temperature susceptibility enhancement.

While several basic ingredients of the model of
Ref.~\onlinecite{vakhitov}
are similar to the assumptions of the present paper, there is an important
distinction: the proposal of
Ref.~\onlinecite{vakhitov}
relies heavily on the bosonization of one-dimensional electrons. As such,
it can be applied to study of Q1D systems, like
o-TaS$_3$
and blue bronze
Rb$_{0.3}$MoO$_3$.
However, non-magnetic tritellurides
LaTe$_3$
and
YTe$_3$
are quasi-two-dimensional.
Thus, it appears important to develop an alternative mechanism, operational
beyond Q1D realm. Our formalism relies not on the bosonization, but rather
on the mean field theory for the Fr\"olich Hamiltonian. The mean field
approach may be used for quasi-two-dimensional, three-dimensional and, with
certain
care~\cite{MFI},
even for Q1D systems. Consequently, our mechanism has a much wider
applicability range.

\section{Conclusions}
%%%%%%%%%%%%%%%%%%%%%%%%%%%%%%%%%%%%%%%%%%%%%%%%%%
\label{sect::conclusions}
%%%%%%%%%%%%%%%%%%%%%%%%%%%%%%%%%%%%%%%%%%%%%%%%%%

In conclusion, we have proposed a possible mechanism responsible for the
generation of the localized magnetic moments in a material with CDW. Its
main idea is quite generic: for an Anderson impurity in an insulating
environment one can always find a parameter range where the impurity hosts
a single electron spin. Using the perturbation theory in the impurity
interaction strength $U$, we mapped the zero-temperature phase diagram of a
single impurity. It consists of magnetic and non-magnetic phases. The
presence of the magnetic phase can affect the thermodynamic properties of a
diluted ensemble of such impurities. It was determined that the ensemble's
susceptibility diverges at low temperature, and the heat capacity
demonstrates marked dependence on weak magnetic field. Both theoretical
findings are consistent with the experimental observations in some CDW
materials.

The mechanism is fairly robust in the sense that the nature of the impurity
and some other details are not very important. While our Hamiltonian
described a point-like impurity, an extended shallow level bound to a
defect may be considered instead. At the same time, the mechanism is not
universal: if the ensemble parameters lie outside the relevant region, no
localized magnetic moments appear, and the material has trivial magnetic
properties.

\section*{Acknowledgments}

This work was partly supported by
the Russian Foundation for Basic Research (projects
Nos.~14-02-00276, 12-02-00339).
The authors would like to thank S.~Artemenko, S.~Zaitsev-Zotov,
D.~Shapiro, and other participants of the condensed matter seminar of
Institute of Radio-engineering and Electronics RAS for useful comments and
suggestions.

\end{document}